\documentclass[12pt]{article}

\usepackage{graphicx}

\def\lsim{\mathrel{\rlap{\lower4pt\hbox{\hskip1pt$\sim$}}
    \raise1pt\hbox{$<$}}}               
\def\gsim{\mathrel{\rlap{\lower4pt\hbox{\hskip1pt$\sim$}}
    \raise1pt\hbox{$>$}}}               

\newcommand{\be}{\begin{eqnarray}}
\newcommand{\ee}{\end{eqnarray}}

\begin{document}

\rightline{\Large{Preprint RM3-TH/03-16}}

\vspace{2cm}

\begin{center}

\LARGE{Possible evidence of extended objects\\[2mm] inside the proton\footnote{\bf To appear in Nucler Physics A as Proceedings of the XVII International IUPAP Conference on {\em Few-Body Problems in Physics}, Durham (North Carolina, USA), June 5-10, 2003.}}

\vspace{2cm}

\large{R.~Petronzio$^*$, S.~Simula$^{**}$ and G.~Ricco$^{***}$}

\vspace{1cm}

\normalsize{$^*$Universit\`a di Roma "Tor Vergata" and INFN, Sezione di Roma II, I-00133 Roma, Italy\\
$^{**}$Istituto Nazionale di Fisica Nucleare, Sezione di Roma III, I-00146 Roma, Italy\\
$^{***}$Universit\'{a} di Genova and INFN - Genova, I-16146, Genova, Italy}

\end{center}

\vspace{2cm}

\begin{abstract}

\noindent Recent data on the Nachtmann moments of the unpolarized proton structure function $F_2^p$, obtained at low momentum transfer with the $CLAS$ detector at Jefferson Lab, are interpreted in terms of the dominance of the elastic coupling of the virtual photon with extended substructures inside the proton. The $CLAS$ data exhibit a new type of scaling behavior and the resulting scaling function can be interpreted as a constituent form factor consistent with the elastic nucleon data. A constituent size of $\approx 0.2 \div 0.3 ~ fm$ is obtained.

\end{abstract}

\newpage

\rightline{}

\newpage

\pagestyle{plain}

\indent The inclusive electron-proton cross section has been recently measured in Hall B at Jefferson Lab using the $CLAS$ spectrometer \cite{CLAS}. The measurements have been performed in the nucleon resonance regions ($W < 2.5 ~ GeV$) for values of the squared four-momentum transfer $Q^2$ below $\approx 4.5 ~ (GeV/c)^2$. One of the most relevant features of such measurements is that the $CLAS$ large acceptance has allowed to determine the cross section in a wide two-dimensional range of values of $Q^2$ and $x = Q^2 / 2M \nu$. This has made it possible to extract the proton structure function $F_2^p(x, Q^2)$ and to directly integrate all the existing world data at fixed $Q^2$ over the whole significant $x$-range for the determination of the (inelastic) transverse Nachtmann moments $M_n^{(T)}(Q^2)$ with order $n = 2, 4, 6, 8$, defined as
 \be
    M_n^{(T)}(Q^2) \equiv \int_0^{x_{\pi}} dx ~ {\xi^{n+1} \over x^3} {3 + 
    3 (n + 1) r + n (n + 2) r^2 \over (n + 2) (n + 3)} ~ F_2^p(x, Q^2)
    \label{eq:MTn} ~,
 \ee
where $\xi = 2 x / (1 + r)$, $r = \sqrt{1 + 4 M^2 x^2 / Q^2}$ and $x_{\pi}$ is the pion threshold.

\indent A possible interpretation of the experimental results of Ref.~\cite{CLAS} has been proposed in Ref.~\cite{PSR}. There the original two-stage model of Ref.~\cite{Altarelli}, developed in the Deep Inelastic Scattering ($DIS$) regime, is extended to values of $Q^2$ around and below the scale of chiral symmetry breaking, $\Lambda_{\chi}$, and above the $QCD$ confinement scale, $\Lambda_{QCD}$, i.e. $0.1 \div 0.2 \lsim Q^2 ~ (GeV/c)^2 \lsim 1 \div 2$. In this contribution we briefly recall the main results of Ref.~\cite{PSR}, namely: ~ i) the data of Ref.~\cite{CLAS} exhibit a new type of scaling behavior expected within the generalized two-stage model, and ~ ii) the resulting scaling function can be interpreted as (the square of) a constituent quark ($CQ$) form factor with a $CQ$ size of $\approx 0.2 \div 0.3 ~ fm$.

\indent The basic assumption of the two-stage model is that hadrons are made of a finite number of $CQ$'s having a partonic structure. The latter depends only on short-distance physics, independent of the particular hadron, while the motion of the $CQ$'s inside the hadron reflects non-perturbative physics depending on the particular hadron. For values of $Q^2$ around and below the scale of chiral symmetry breaking, one expects that the {\em elastic} coupling of the incoming virtual boson with the $CQ$ dominates over the {\em inelastic} channels. As explained in Ref.~\cite{PSR}, the $Q^2$-range of applicability of the generalized two-stage model is qualitatively given by $\Lambda_{QCD}^2 \lsim Q^2 \lsim \Lambda_{\chi}^2$, i.e. $0.1 \div 0.2 \lsim Q^2 ~ (GeV/c)^2 \lsim 1 \div 2$. 

\indent The dominance of the elastic coupling at the $CQ$ level cannot however hold at each $x$ value, but only in a local duality sense. The $CQ$-hadron duality can be translated in the space of moments into the following (approximate) equivalence
 \be
    M_n^{(T)}(Q^2) \simeq F^2(Q^2) \cdot \overline{M}_n^{(T)}(Q^2) ~,
    \label{eq:dualT}
 \ee
where  $\overline{M}_n^{(T)}(Q^2)$ describes the effect of the internal $CQ$ motion inside the hadron on the moment of order $n$, and $F(Q^2)$ is the $CQ$ elastic form factor (see Ref.~\cite{PSR} for its definition). Equation (\ref{eq:dualT}) is expected to hold for low values of the order $n$, except $n = 2$, because the second moment $M_2^{(T)}(Q^2)$ is significantly affected by the low-$x$ region where the concept of valence dominance may become unreliable (cf. for more details Ref.~\cite{PSR}).

\indent If one has a reasonable model for the $CQ$ momentum distributions in the hadron, the theoretical moments $\overline{M}_n^{(T)}(Q^2)$ can be estimated and therefore the ratio
 \be
    R_n^{(T)}(Q^2) \equiv M_n^{(T)}(Q^2) ~ / ~ \overline{M}_n^{(T)}(Q^2)
    \label{eq:ratioT}
 \ee
can be constructed starting from the experimental moments $M_n^{(T)}(Q^2)$. Thus, if Eq.~(\ref{eq:dualT}) holds, the ratio $R_n^{(T)}(Q^2)$ is expected to depend only on $Q^2$, i.e. it becomes independent of the order $n$ (as well as on the specific hadron), viz. 
 \be
    R_n^{(T)}(Q^2) \simeq F^2(Q^2) ~.
    \label{eq:scaling}
 \ee
The scaling function, given by the r.h.s. of Eq.~(\ref{eq:scaling}), is directly the square of the $CQ$ form factor, i.e. the form factor of a confined object.

\indent The data of Ref.~\cite{CLAS} manifest a clear tendency to the scaling property (\ref{eq:scaling}) even assuming no internal $CQ$ motion inside the proton, which represents a very simplified and rough model for the $CQ$ momentum distribution. Explicitly one gets $\overline{M}_n^{(T)}(Q^2) \to (1/3)^{n-1}$. Though simple, such an hypothesis explains very well the spread of about one order of magnitude between the experimental moments of order $n$ and $(n + 2)$ (cf. Ref.~\cite{PSR}). In our opinion this is an important result (almost a pure experimental result) because obtained with a very simple hypothesis about the $CQ$ motion in the proton. 

\begin{figure}[htb]

\centerline{\includegraphics[scale=0.7]{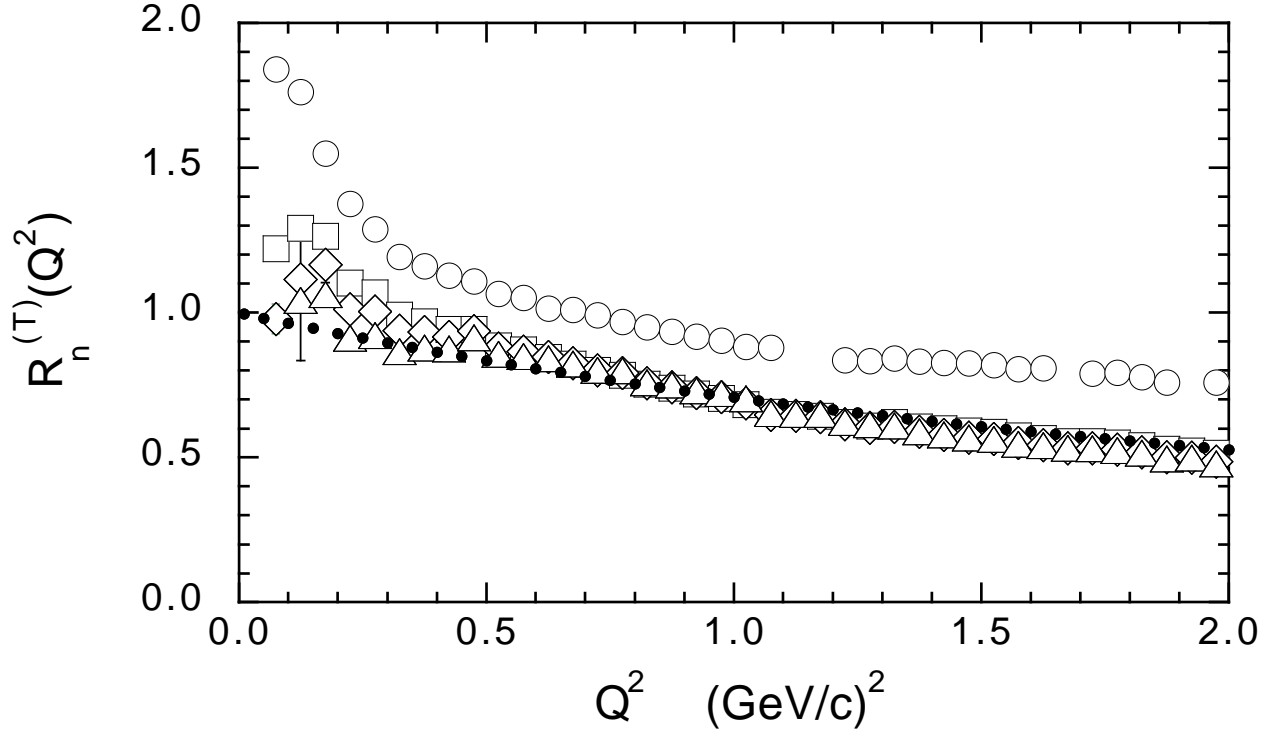}}

\caption{\label{fig:scaling} \small Ratio $R_n^{(T)}(Q^2)$ of the experimental moments $M_n^{(T)}(Q^2)$, obtained in Ref.~\cite{CLAS}, with the theoretical moments $\overline{M}_n^{(T)}(Q^2)$, calculated in Ref.~\cite{PSR}. The dotted line represents the square of a monopole form factor corresponding to a $CQ$ size $r_Q = 0.21 ~ fm$. The dots, squares, diamonds and triangles correspond to $n = 2, 4, 6$ and $8$, respectively. (Adapted from Ref.~\cite{PSR}).}

\end{figure}

\indent The effect of the internal $CQ$ motion was investigated in Ref.~\cite{PSR} and found to be important. The basic result of Ref.~\cite{PSR} is represented in Fig.~\ref{fig:scaling}, where it can clearly be seen that the scaling property (\ref{eq:scaling}) is well satisfied by the $CLAS$ data for $n = 4, 6, 8$ with the expected exception of $n = 2$. The scaling function of Fig.~\ref{fig:scaling} can be interpolated using the square of a monopole ansatz, $F(Q^2) \simeq 1 / (1 + r_Q^2 ~ Q^2 / 6)$, with $r_Q = 0.21 ~ fm$. The precise value of the $CQ$ size $r_Q$ depends on the specific values of the model parameters. Such a dependence has been thoroughly investigated in Ref.~\cite{PSR} and the final result is that a safe estimate of the $CQ$ size $r_Q$ is between $\approx 0.2$ and $\approx 0.3 ~ fm$.

\indent The $CQ$ form factor extracted from the scaling function and the model used for the nucleon wave function should be consistent with the elastic nucleon data. In Ref.~\cite{PSR} the nucleon elastic form factors were calculated adopting the covariant light-front approach of Ref.~\cite{nucleon}, which is properly formulated at $q^+ = 0$. It turns out that the calculated nucleon form factors slightly overestimate the data, but a nice consistency can be easily reached through small variations of the model parameters.

\indent In conclusion the generalized two-stage model of Ref.~\cite{PSR} provides an explanation of the recent $CLAS$ data of Ref.~\cite{CLAS} in terms of the dominance of the elastic coupling of the virtual photon with extended objects inside the proton at low values of $Q^2$, namely $0.1 \div 0.2 \lsim Q^2 ~ (GeV/c)^2 \lsim 1 \div 2$. A positive comparison with forthcoming data on other structure functions, like the longitudinal (see Ref.~\cite{longitudinal}) and the polarized ones, will provide further compelling evidence that constituent quarks are intermediate substructures between the hadrons and the current quarks and gluons of $QCD$.

\end{document}